\documentclass[twocolumn,showpacs,preprintnumbers,amsmath,amssymb]{revtex4}

\usepackage{graphicx}
\usepackage{dcolumn}
\usepackage{bm}
\usepackage{color}

\begin{document}

\preprint{APS/123-QED}

\title{ Improved sensitivity of magnetic measurements under high pressure in miniature ceramic anvil cell for a commercial SQUID magnetometer\footnote{Rev. Sci. Instrum. {\bf 84}, 046105 (2013)}}

\author{Naoyuki Tateiwa$^{1}$}
\author{Yoshinori Haga$^{1}$}%
\author{Tatsuma D Matsuda$^{1}$}%
\author{Zachary Fisk$^{1,2}$}
\author{Shugo Ikeda$^{3}$}
\author{Hisao Kobayashi$^{3}$}

\affiliation{
$^{1}$Advanced Science Research Center, Japan Atomic Energy Agency, Tokai, Naka, Ibaraki 319-1195, Japan\\
$^{2}$University of California, Irvine, California 92697, USA\\
$^{3}$Graduate School of Material Science, University of Hyogo, Koto, Hyogo 678-1297, Japan\\
}
\date{\today}

\begin{abstract}
   Two modifications have been made to a miniature ceramic anvil high pressure cell (mCAC) designed for magnetic measurements at pressures up to 12.6 GPa in a commercial superconducting quantum interference (SQUID) magnetometer [N. Tateiwa {\it et al.,} Rev. Sci. Instrum. {\bf 82}, 053906 (2011)., ibid. {\bf 83}, 053906 (2012)]. Replacing the Cu-Be piston in the former mCAC with a composite piston composed of the Cu-Be and ceramic cylinders reduces the background magnetization significantly smaller at low temperatures, enabling more precise magnetic measurements at low temperatures. A second modification to the mCAC is the utilization of a ceramic anvil with a hollow in the center of the culet surface. High pressures up to 5 GPa were generated with the ``cupped ceramic anvil" with the culet size of 1.0 mm.
\end{abstract}

\maketitle

 Magnetic measurement at high pressure is an important experimental method for the study of magnetic properties of materials at high pressure. Recently, we have proposed a miniature ceramic-anvil high-pressure cell for magnetic measurements at pressures above 10 GPa in a commercial superconducting quantum interference (SQUID) magnetometer \cite{tateiwa1,tateiwa2,tateiwa3}. This cell is abbreviated mCAC. Readers can refer to the ref. 1 and 2 for details of the mCAC and the current status of the pressure cells developed for the SQUID magnetometer. The mCAC has several advantages as described in the references. The anvils are made of inexpensive composite ceramic (FCY20A, Fuji Die Co.). A problem in the mCAC is that although the background magnetization is small at higher temperatures, the absolute value increases with decreasing temperature below 10 K, which is disadvantageous for the high pressure study in the low temperature physics.  In this paper, we report two modifications in the mCAC: one is the reduction of background magnetization at low temperatures and the other is the application of the ``cupped anvil" for generating a more hydrostatic pressure.  
  
    \begin{table}[b]
\caption{\label{tab:table1}%
Information on the Cu-Be rods. JIS: Japanese Industrial Standards. UNS: Unified Numbering System.}
\begin{ruledtabular}
\begin{tabular}{cccccccc}
\textrm{rod No.}&
\textrm{JIS No.  (UNS)}&
\textrm{Company}\\
\colrule
No. 1 &C1720 (C17200) &Yamatogokin Co., Ltd.\\
2 &C1720 (C17200) &Yamatogokin Co., Ltd.\\
3 &C1720 (C17200)  &Nilaco Corporation \\
4 & C1720 (C17200)  &Nilaco Corporation\\
5 & C1720 (C17200)  &Materion Brush Japan, Ltd. \\
6 & C1715 (C17150)  &Materion Brush Japan, Ltd. \\
\end{tabular}
\end{ruledtabular}
 \end{table}

    \begin{figure}[]
\includegraphics[width=8cm]{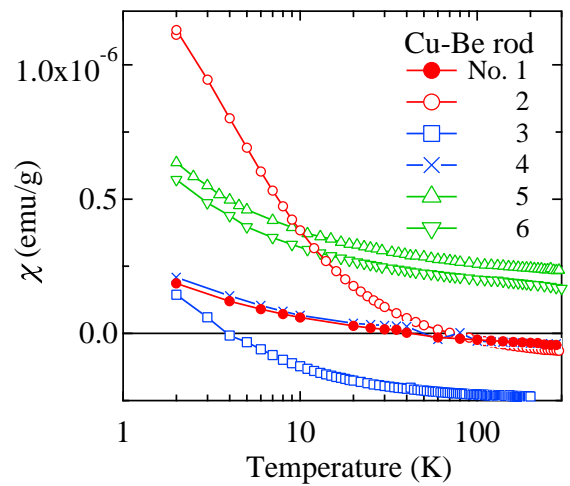}
\caption{\label{fig:epsart}(Color online) Temperature dependences of the magnetic susceptibility $\chi$ in the Cu-Be alloys (rods).}
\end{figure} 
   \begin{figure}[]
 \includegraphics[width=8cm]{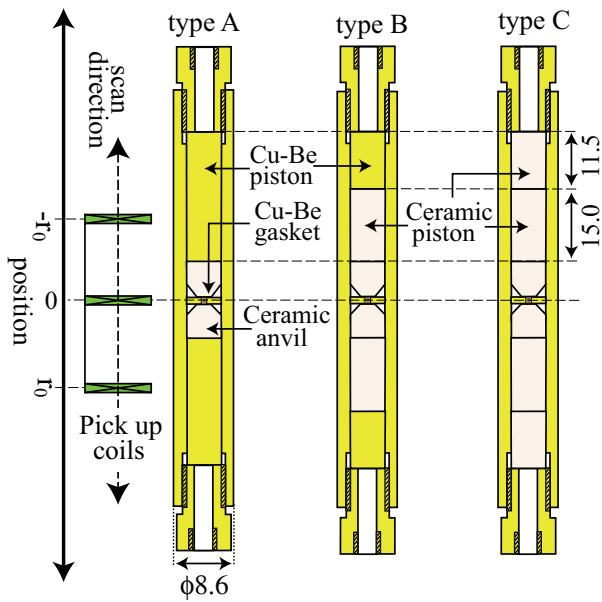}
\caption{\label{fig:epsart}(Color online) Cross-sectional views of three types (type A, B, and C) of the miniature high pressure cell (mCAC) for the commercial SQUID magnetometer.  }
\end{figure} 

   Figure 1 shows the temperature dependences of the magnetic susceptibility $\chi$ in magnetic field of 10 kOe in the standard Cu-Be C1720 (JIS: Japanese Industrial Standards) and C1715 alloys (rods) purchased from several companies as shown in Table. I. The temperature dependence of $\chi$ consists of an almost temperature independent diamagnetism from the alloy and a paramagnetic Curie-Weiss term from Ni or Co impurities.  The behavior of $\chi$ depends on each rod.  In our study, all pressure cells and gaskets were made of the Cu-Be rod No. 1.
         \begin{figure}[]
\includegraphics[width=8cm]{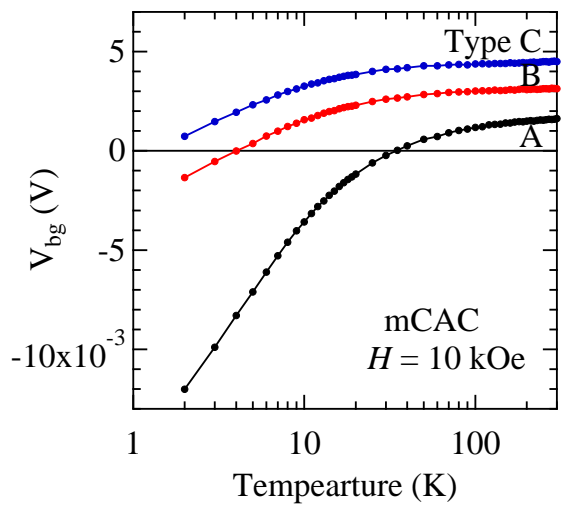}
\caption{\label{fig:epsart}(Color online)Temperature dependences of the back ground magnetization $V_{bg}$(V) in the three types of the mCAC in magnetic field of 10 kOe. }
\end{figure}

 Figure 2 shows several designs of the mCAC for a magnetic property measurement system  (MPMS) from Quantum Design (USA). Type A was previously reported by us\cite{tateiwa1,tateiwa2,tateiwa3}. Type B and C are newly designed in this study. The Cu-Be piston in the type A is replaced with the ceramic and Cu-Be cylinders in the type B, and the two ceramic cylinders in the type C. In the SQUID magnetometer, the cell moves through three pick up coils which detect inductive voltage. MPMS has three pick up coils and two coils are placed with the distance of $r_0$ = 15.2 mm from the center coil as shown in Figure 2\cite{mpms}. We compare the background magnetization from  the difference between the scaled voltage inducted in the coil at $r=0$ and the one at $r=r_0$: $V_{bg}$ = $V(0)$ - $V({r_0})$. We note that $V(0)$ - $V({r_0})$ is roughly proportional to the magnetization of a sample.

 Figure 3 shows the temperature dependences of the background magnetization $V_{bg}$(V) in the mCAC of the three types (A, B, and C) in magnetic field of 10 kOe. The voltage $V_{bg}$ of the former type A is smaller than those of the new types B and C above 20 K. It decreases with decreasing temperature and changes its sign near 30 K. The absolute value of $V_{bg}$ at the lowest temperature 2 K is about seven times larger than that at room temperature. This large background magnetization is a problem in the type A mCAC as mentioned in the introduction. Meanwhile, $V_{bg}$ in types B and C does not show a strong temperature dependence. The absolute values of $V_{bg}$ are about 10 \% of that of type A at 2.0 K. For magnetic measurements at higher temperatures, the type A mCAC should be used, while the new types B and C are appropriate for the experiments at low temperatures.  The cell structure around the upper and lower coils has a strong influence on the background magnetization of the cell. The background magnetization of the type B and C at 2.0 K are roughly estimated as $-3$ $\times$ 10$^{-4}$ and 2 $\times$ 10$^{-4}$ emu, less than 5 $\%$ of that in the indenter cell\cite{kobayashi}.

 Next we show the other modification: test of a ceramic anvil which has a hollow at the center of the culet surface. This anvil is abbreviated here as  ``cupped ceramic anvil". The cupped anvil, originally designed by Ivanov and Vereshchagin in 1960, has been used in high pressure studies for a long time \cite{eremets}. The cupped anvil provides a larger sample space and makes possible smaller pressure distribution at the culet surface of the anvil\cite{eremets}. The cupped anvil has been made of metallic WC or sintered diamond. We show the utility of the cupped structure in the anvil made of the ceramic material. 
    \begin{figure}[t]
\includegraphics[width=8cm]{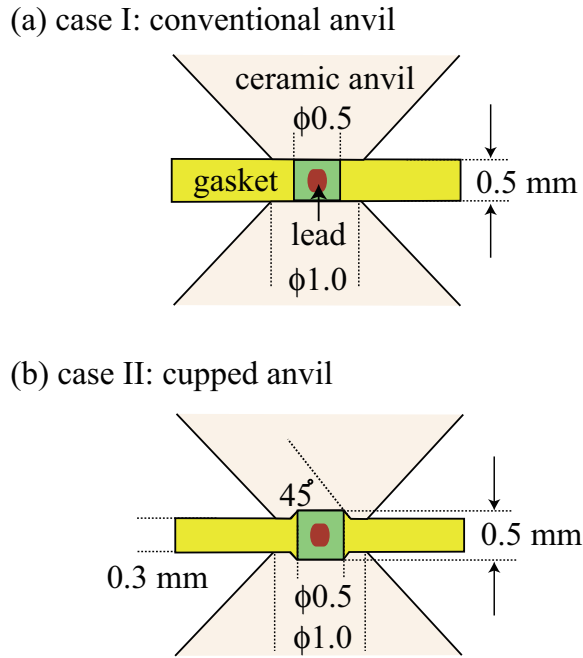}
\caption{\label{fig:epsart}(Color online)Schematic illustrations of cross-section of the conventional anvil without cupped structure and (b) ``cupped ceramic anvil" with the Cu-Be gasket. }
\end{figure}

 We compare the performance of the conventional and the cupped ceramic anvils with the culet size of 1.0 mm. Figure 4 shows schematic illustrations of cross-section of (a) (case I) the conventional anvil without the cupped structure and the Cu-Be gasket with the thickness = 0.5 mm, and (b) (case II) the cupped ceramic anvil and the gasket with the thickness = 0.3 mm. The diameter and depth of the concave shape in the cupped anvil are 0.7 and 0.1 mm, respectively. The volume of the sample space for case I and II are almost the same. We have studied the relations between the applied load at room temperature and the pressure value at low temperatures in both cases. The sample and the lead (Pb) pressure manometer were placed in the sample space filled with the pressure-transmitting medium glycerin\cite{tateiwa4}. The pressure values at low temperatures were determined by the pressure dependence of the superconducting transition temperature in lead\cite{wittig}.

 Figure 5 (a) shows the applied load dependences of the pressure value at low temperatures for the case I and II. In the case I, the pressure value increases with increasing load and starts to saturate at higher applied load. The maximum pressure is about 4.2 GPa. Meanwhile, the pressure increases linearly with increasing pressure in the case II and the pressure efficiency is larger than that in the former case. The maximum pressure is 5.0 GPa.  The initial volume of the sample space in both cases are almost similar. 
  \begin{figure}[b]
\includegraphics[width=8cm]{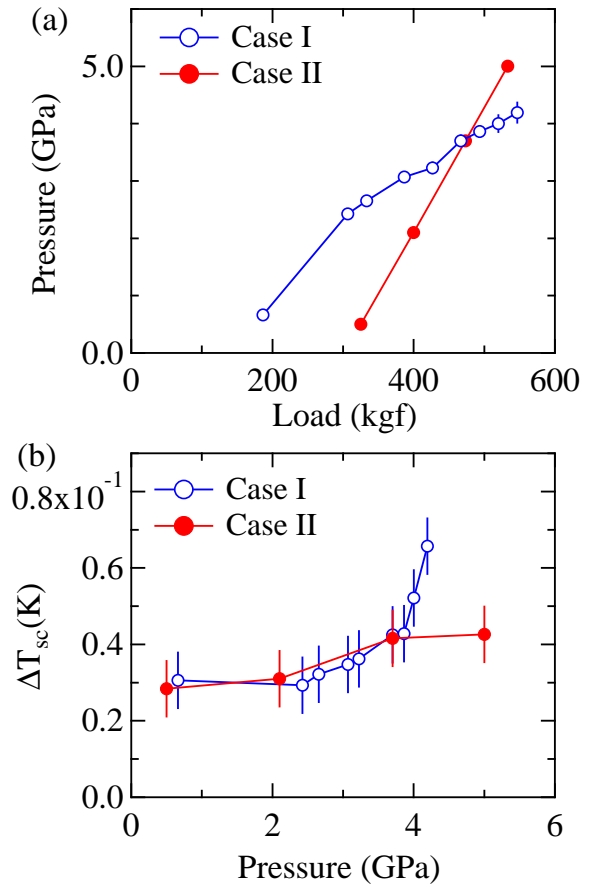}
\caption{\label{fig:epsart}(Color online) (a)Applied load dependences of the pressure value at low temperatures for the case I and II. (b)Pressure dependences of the width of the superconducting transition ${\Delta}T_{sc}$ in lead for both cases.}
\end{figure} 

The superconducting transition of the lead becomes broadened when the pressure inside the sample space deviates from hydrostatic. Figure 5 (b) shows pressure dependences of ${\Delta}T_{sc}$ in lead for the case I and II. The value of ${\Delta}T_{sc}$ starts to increase largely above 4 GPa in the case I, suggesting deviation from the hydrostatic pressure above this pressure. The initial value of the diameter of the sample space was 0.5 mm. After the experiment of case I, the diameter of the sample space was increased to 0.55 mm, while the thickness decreased. This radial deformation leads to the development of uniaxial stress in the sample space and the pressure deviates from hydrostatic.  Meanwhile, the diameter in case II decreased from the initial value after the experiment, indicating that the sample space was compressed more isotropically. There is no clear increase in the pressure dependence of ${\Delta}T_{sc}$, suggesting the absence of a strong deviation from hydrostatic pressure. 

   In summary, we made two modifications to a miniature ceramic anvil high pressure cell (mCAC) designed previously by us. The background magnetization becomes very small at low temperatures and the value at 2.0 K is about 10  \% of that of the previous version. This modification enables more precise magnetic measurements at high pressure and at low temperatures. The other modification in the mCAC is the introduction of a ceramic anvil that has a hollow in the center of the culet surface.  High pressures up to 5 GPa were generated with the ``cupped ceramic anvil" with the culet size of 1.0 mm.

 We thank Dr. Keiki Takeda (Muroran Institute of Technology, Japan) for the critical reading of this manuscript and enlighten suggestions on the high pressure technology for a long time. This work was supported by a Grant-in-Aid for Scientific Research on Innovative Areas ``Heavy Electrons (Nos. 20102002 and 23102726), Scientific Research S (No. 20224015), A(No. 23246174), C (No. 22540378) from the Ministry of Education, Culture, Sports, Science and Technology (MEXT) and Japan Society of the Promotion of Science (JSPS).

\bibliography{apssamp}

\begin{references}



\bibitem{tateiwa1}N. Tateiwa, Y. Haga, Z. Fisk, and Y. {\=O}nuki, Rev. Sci. Instrum. {\bf 82}, 053906 (2011).

\bibitem{tateiwa2}N. Tateiwa, Y. Haga, and Z. Fisk,  Rev. Sci. Instrum. {\bf 83}, 053906 (2012).

\bibitem{tateiwa3}N. Tateiwa, and Y. Haga, Japanese Patent Tokugan No. 2011-054153 (pending).




\bibitem{mpms}Quantum Design Co., web site: http://www.qdusa.com/

\bibitem{kobayashi}T. C. Kobayashi, H. Hidaka, H. Kotegawa, K. Fujiwara, and M. I. Eremets, Rev. Sci. Instrum. {\bf 78}, 023909 (2007).

\bibitem{eremets}M. I. Eremets,  {\it High-pressure Experimental Methods} (Oxford University Press, Oxford, 1996). 

\bibitem{tateiwa4}N. Tateiwa and Y. Haga,  Rev. Sci. Instrum. {\bf 80}, 123901 (2009).

\bibitem{wittig} B. Bireckoven and J. Wittig, J. Phys. E: Sci. Instrum.  {\bf 21}, 841 (1988).



\end{references}

 \newpage

\end{document}